\documentstyle[12pt]{article}

\begin{document}
\baselineskip 18pt

\begin{center}
{\Large\bf   Dynamic nonlinear (cubic) susceptibility

 in quantum Ising spin glass}\\

\bigskip

{G. Busiello$^a$\footnote{Corresponding
author,tel.(0)39-89-965426 -
fax (0)39-89-965275 - e-mail: busiello@sa.infn.it}}\\

\medskip

{\small\it Dipartimento di  Fisica ``E.R. Caianiello",
Universit\`a
di Salerno,\\
84081 Baronissi - Salerno and INFM - Unit\`a  di Salerno, Salerno, Italy}\\
\medskip
{R. V. Saburova and V. G. Sushkova}\\
{\small\it Kazan State Power-Engineering University, Kazan, Russia}
\end{center}

\bigskip
{\begin{abstract}   Dynamic nonlinear (cubic) susceptibility in
quantum d-dimensional Ising spin glass with short-range
interactions is investigated on the basis of quantum droplet model
and quantum-mechanical nonlinear response theory. Nonlinear
response depends on the tunneling rate for a droplet which
regulates the strength of quantum fluctuations.  It shows a strong
dependence on the distribution of droplet free energies and on the
droplet length scale average. Comparison with recent experiments
on quantum spin glasses like disordered dipolar quantum Ising
magnet is discussed .\end{abstract}}

\bigskip \bigskip \bigskip \bigskip \bigskip

\newpage

The dynamics of glassy systems is an attractive and rapidly
developing field of physics [1-5]. Spin glasses and quantum spin
glasses are very interesting systems for theoretical and
experimental investigation of dynamic phenomena [2-4]. Two major
theoretical description of spin glass phenomena have evolved over
the past twenty years:the mean field theory based on the Parisi
replica symmetry-breaking approach and the droplet model based on
renormalization group arguments [1-6]. The two pictures give very
different physical interpretation of observable spin-glass
phenomena. In this paper we investigate theoretically nonlinear
cubic dynamic susceptibility as function of frequency and
temperature in the Ising spin glass in a transverse field in terms
of quantum droplet model at very low temperatures (quantum
regime). The quantum phase transition is governed by quantum
fluctuations of the system which may tunnel from one local minimum
of the free energy to another; new physical effects such as
quantum channel of relaxation appear. There are few theoretical
studies on the nonlinear static response in quantum spin glasses
[7-10] and almost no studies on the dynamic nonlinear response
[3]. In [3] dynamic nonlinear response of a quantum spin glass was
found frequency independent and nonsingular in quantum critical
regime in contrast to the behavior in usual spin glass. There are
experimental data on the nonlinear dynamic response in classical
[11-13] and quantum [14] spin glasses
 investigated by Fourier-transform technique. The
third-order nonlinear susceptibility is negative and diverges at
an ordinary spin glass transition temperature $T_{g}$ from both
the upper and the lower sides. But when $\chi'_{3}$ is measured by
a finite probing frequency the response falls out equilibrium
before the transition temperature and does not diverge at $T_{g}$.
Then $\chi'_{3}(\omega)$ shows a maximum at $T\simeq
T_{f}(\omega)$ where $T_{f}(\omega)$ is the freezing temperature
which is the upper bound on $T_{g}$ and $T_{g}=T_{f}(\omega
\rightarrow 0)$. Such behavior was observed, for example, for
classical Ising spin glass ${\rm Fe}_{0.5}{\rm Mr}_{0.5}{\rm
TiO}_{3}$ [13]. W. Wu et al. [14] measured nonlinear
susceptibility $\chi'_{3}(\omega , T)$ in quantum spin glass (the
diluted dipolar-coupled Ising spin glass ${\rm LiHo}_{0.167}{\rm
Y}_{0.833}{\rm F}_{4}$ in the transverse field) tuning transverse
field $\Gamma$ from the $\Gamma = 0$ classical to the $T=0$
quantum limit. At $mK$ temperatures they found a clear dynamic
signature of the spin glass to paramagnet transition whether
dominated by thermal or quantum fluctuations. In [14] it was shown
that the $\chi'_{3}$ depends on frequency for $\omega > 10 Hz$.
However it depends very weakly on $\omega$ for $\omega < 10Hz$.
There is a crossover between high $\omega$ ($\omega$ - dependent)
and low $\omega$ ($\omega$ - independent) behaviors. Nonlinear
susceptibility contains a diverging component which dominates at
$T=98 mK$, but disappears by $25 mK$. The $\chi'_{3}(\omega)$ does
not diverge but shows a maximum at $T_{f}(\omega)$.
$\chi'_{3}(\omega)$ measured at a higher temperature and lower
transverse field has a larger maximum than $\chi'_{3}(\omega)$
measured at a lower temperature and larger transverse field. The
analysis of these experimental data seems not clear [15] because
frequencies used in the experiments [14] are not sufficiently low
such as to determine the equilibrium behavior of system. Contrary
to the theoretical expectations, quantum transitions may be
qualitatively different from thermally driven transitions in real
spin glasses. Recently the linear dynamic susceptibility (in-phase
and out of phase components) at $T=0$ was investigated
theoretically for the Ising spin glass in a transverse field in
terms of quantum droplet model by M.~J.~Thill and D.~A.~Huse [6].
 A basic assumption of the droplet picture is that the spin glass
dynamics is governed by large-scale excitations whose relaxation
time increases with length scale. In previous papers [16] we have
calculated the real and imaginary parts of linear dynamic
susceptibility in the same model, as in [6], for very low nonzero
temperatures using the general linear response theory of magnetic
dispersion and absorption phenomena for quantum systems by Kubo
and Tomita [17]. We note that in [6] the real part of the cubic
nonlinear ac susceptibility was defined as the in-phase $3\omega$
magnetization response $M(3\omega)$ to a small time-dependent
applied field $h\cos (\omega t)$
\begin{equation}
\chi'_{3}=\lim_{h\rightarrow 0}{24M(3\omega)\over h^{^3}V}
\end{equation}
where $V$ is the sample volume. The authors of [6] gave some
expression for $\chi'_{3}$ they only expect at zero temperature.

The full nonlinear response theory despite its generality and
importance is of limited practical value because it is
mathematically difficult. It is necessary to make approximations,
as the well-known perturbation expansion of the time-evolution
operator, using  Hamiltonian with some small parameter. Nonlinear
response theory was developed and described, for example, in
[17-22]. Here we summarize briefly the theory of higher-order
dynamic response. It is based on the Hamiltonian
\begin{equation}
\hat{\cal H}_{t}=\hat{\cal H}_{0}+\hat{\cal H}_{1}=\hat{\cal
H}_{0}-\hat{A}_{j}F_{j}(t),\,\,t \geq t_{0}
\end{equation}
where $\hat{\cal H}_{0}$ is nonperturbation Hamiltonian of system,
$\hat{\cal H}_{1}$ is perturbation Hamiltonian which describes the
interaction between the Heisenberg operators $\hat{A}_{j}$
(material operators) and the external perturbation $\hat{F}_{j}$.
At time $t>t_{0}$ one is interested in the expectation value of
the Heisenberg operator $\hat{B}_{i}$ which is given by
\begin{equation}             
\langle\hat{B}_{i}(t)\rangle = Tr[\rho_{0}\hat{B}_{i}(t)] =
Tr[\hat{\rho}(t)\hat{B}_{i}]
\end{equation}
where the density matrix $\hat{\rho}(t)=\hat{U}(t,
t_{0})\hat{\rho}(t_{0})\hat{U}^{\dagger}(t,t_{0})$. The
time-evolution operator $\hat{U}$ satisfies the Schr\"odinger
equation $ih{d\hat{U} \over dt}=\hat{\cal H}\hat{U}$.  It is
difficult to find an expression for $\hat{U}$ in closed form. It
was used that $\hat{\cal H}_{1}(t)$ is in some sense small. We
define the total response of the system at time $t$ to the
external force $F_{j}$ as the difference
\begin{equation}   
\Delta B_{i}(t)\equiv \langle\hat{B}_{i}(t)\rangle-\langle\hat{B}_{i}\rangle_{0}
\end{equation}
where the subscript zero on expectation values refers to the
equilibrium expectations. One can then understand the behavior of
the system in terms of the dynamical response. Using
aforementioned expressions [2-3] the dynamical response can be
written through the third order in the perturbation $\hat{\cal
H}_{1}$ in the following form [20]
$$
  \bigl<\hat{B}_{i}(t)\bigr>-\bigl<\hat{B}_{i}\bigr>_{0}=\Delta B_{i}(t)\simeq
  \int_{t_0}^tdt'\varphi_{ij}(t-t^{'})F_{j}(t^{'})+ $$
$$
 \int_{t_{0}}^{t}dt_{1}
\int_{t_{0}}^{t_1}dt_{2}\,\varphi_{ijk}(t-t_{2},
t_{1}-t_{2})F_{k}(t_{1})F_{j}(t_{2}) +  $$
\begin{equation}  \int_{t_{0}}^{t}dt_1
\int_{t_{0}}^{t_1}dt_2\int_{t_{0}}^{t_2}dt_3\,\varphi_{ijkl}(t-t_{3}, t_{2}-t_{3},
t_{1}-t_{2})F_{e}(t_{1})F_{k}(t_{2})F_{j}(t_{3})+ \ldots
\end{equation}
where $\varphi_{ij}$, $\varphi_{ijk}$, $\varphi_{ijkl}$ are the
first-, second- and third order response functions,
\begin{equation}   
\varphi_{ij}(t-t^{'})={1 \over ih}\langle[\hat{A_{j}},
B_{i}(t-t^{'})]\rangle_{0},
\end{equation}

\begin{equation}   
\varphi_{ijk}(t-t_{2}, t_{1}-t_{2}) = {1\over
(ih)^{2}}\langle[\hat{A}_{j}, [\hat{A}_{k} (t_{1}-t_{2}),
\hat{B}_{i}(t-t_{2})]]\rangle_{0},
\end{equation}

\begin{equation}   
\varphi_{ijkl}(t-t_{3}, t_{2}-t_{3}, t_{1}-t_{2}) = {1\over
(ih)^{3}} \langle[\hat{A}_{j}, [A_{k}(t_{1}-t_{2}), [A_{l}(t_{2}-t_{3}),
B_{i}(t-t_{3})]]]\rangle_{0}.
\end{equation}
Here we have employed the summation convention over repeated
indices and cyclic invariance of the trace. The expressions (6-8)
can be written in a more revealing form if we set $t_{0}=-\infty$
and change integration variables including an adiabatic switching
factor if necessary. We take (instead of $t$, $t_{1}$, $t_{2}$,
$t_{3}$) $\tau_{1}=t-t_{1}$, $\tau_{2}=t_{1}-t_{2}$ time
differences further. Ordinary linear response theory utilized only
$\varphi_{ij}(\tau_{1})$ and it is the simplest approximation to
the full theory of linear dynamic response [23]. Using new
variables we may write the expressions for response functions in
the form

\begin{equation}   
\varphi_{ij}^{(1)}=-{1 \over ih}\langle[A(\tau), B(0)]\rangle,
\end{equation}

\begin{equation}   
\varphi_{ijk}^{(2)}= {1\over (ih)^{2}}\langle[[A(\tau_{1}+\tau_{2}),
A(\tau_{2})], B]\rangle,
\end{equation}

\begin{equation}   
\varphi_{ijkl}^{(3)}=-{1\over (ih)^{3}}
\langle[[[A(\tau_{1}+\tau_{2}+\tau_{3}), A(\tau_{2}+\tau_{3})],
A(\tau_{3})], B]\rangle
\end{equation}
where the bracket $\langle\ldots \rangle$ denotes an expectation
value with respect to the equilibrium ensemble. The useful
interpretation is generated from eq. (5) in the case that $t_{0}=
-\infty$ if we consider the external force $F$ is constant and
vanishes for $t \geq 0$. For $t=0$ the system is in partial
equilibrium and starts to relax to equilibrium. It is convenient
to write nonlinear response for this case (initial value case
[21]) as

\begin{equation}   
\langle\hat{B}(t)\rangle-\langle\hat{B}\rangle_{0}= R^{(1)}(t)F+{1\over 2}R^{(2)}(t)
FF+{1\over 3}R^{(3)}(t)FFF+\ldots
\end{equation}
where $R^{\alpha}(t)$ are the relaxation functions,
\begin{equation}   
R^{(1)}(t)=\int_{0}^{\infty}d\tau\,\varphi_{ij}^{(1)}(\tau_{1}),
\end{equation}
\begin{equation}            
R^{(2)}(t)=\int_{0}^{\infty}d\tau_{1}
\int_{0}^{\infty}d\tau_{2}\,\varphi_{ijk}^{(2)}(\tau_{2},
\tau_{1}+\tau_{2}),
\end{equation}
\begin{equation}   
R^{(3)}(t)=\int_{0}^{\infty}d\tau_{1}\int_{0}^{\infty}d\tau_{2}
\int_{0}^{\infty}d\tau_{3}\,\varphi_{ijkl}^{(3)}(\tau_{3},
\tau_{2}+\tau_{3}, \tau_{1}+\tau_{2}+\tau_{3}).
\end{equation}

In this form response may describe relaxation of the system. If
response function $\varphi(t)^{(1)}$ vanishes as $t\rightarrow
\infty$, then $\varphi(t)^{(1)}=-{\delta \over \delta t
}R^{(1)}(t)$, so $R^{(1)}(t)$ contains much more information than
the response function.

Let the $ac$ magnetic field $h_{\omega}=h\cos (\omega t)$ be
applied to a magnetic system. The magnetization nonlinear
response $M(\omega, t)=\sum_{k=1}^{n} \{ \Theta_{k}^{'}\cos
(k\omega t)+\Theta^{''}_{k}\sin (k\omega t) \}$ to harmonic
magnetic field $h$ contains only odd harmonic;
$\Theta_{1}^{'}\sim\chi_{1}^{'}h$, $ \Theta_{3}^{'}\sim
\chi_{3}^{'}h^{3}$, etc [11]. In expression for $M(\omega ,
t)$ the magnitudes $\Theta'_k$ and $\Theta''_k$ are
real and imaginary parts of harmonic amplitudes respectively. For the
general theory of nonlinear processes one can evaluate [11]
$$
\Theta_{1}^{'}=\chi_{1}^{'}(\omega t)h + [\chi_{3}^{'}(\omega,
0, \omega)]{h^{3}\over 4}+[4\chi_{5}^{'}(\omega, 0,\omega, 0,
\omega)+2\chi_{5}^{'}(\omega,
0, \omega, 2\omega, \omega) +  $$
\begin{equation}
2\chi_{5}^{'}(\omega, 2\omega, \omega, 0, \omega) +
\chi_{5}^{'}(\omega, 2\omega, \omega, 2\omega, \omega) +
\chi_{5}^{'}(\omega, 2\omega, 3\omega, 2\omega,
\omega)]{h^{5}\over 16}+ \ldots
\end{equation}
$$
\Theta_{3}^{'}=\chi_{3}^{'}(3\omega, 2\omega,
\omega){h^{3}\over 4} + [\chi_{5}^{'}(3\omega, 2\omega, 3\omega,
2\omega, \omega)+\chi_{5}^{'}(3\omega, 4\omega, 3\omega, 2\omega,
\omega) $$
\begin{equation}2\chi_{5}^{'}(3\omega, 2\omega, \omega, 0, \omega) +
\chi_{5}^{'}(3\omega, 2\omega, \omega, 2\omega,
\omega)]{h^{5}\over 16}+ \ldots ,
\end{equation}
\begin{equation}
\Theta_{5}^{'}=\chi_{5}^{'}(5\omega, 4\omega, 3\omega, 2\omega,
\omega){h^{5}\over 16}+ \ldots
\end{equation}
The measurement of all the harmonic amplitudes $\Theta_{k}$ gives
a measurement of the susceptibilities $\chi_{a}$ in two limits: a)
if $\chi_{1}^{'}h\gg \chi_{3}^{'}h^{3} \gg \chi_{5}^{'}h^{5}$ the
back reaction is negligible and each harmonic measures the
susceptibility of the same order; b) in the static ($\omega
\rightarrow 0$) limit the solution of the linear system [15] fully
accounts for the back reaction. In the absence of $ac$ magnetic
field, the back reaction can be made small so that the dynamic
susceptibilities can be obtained from eq. (15). In a more compact
notation we may write
\begin{equation}
M(\omega, t) \sim [M_{0}+M_{\omega}+M_{3\omega}+\ldots]+(complex
\ \  conjugate)
\end{equation}                 
where $M_0$ is the equilibrium magnetization in zero field;
$M_\omega$ is the $\omega $-magnetization response; $M_{3\omega}$
is the $3\omega$-magnetization response and so on.

The expressions (9-11) may be considered as solution of the
corresponding quantum equations considered above. The external
$ac$ field we assume classical value. This field interacts with
quantum system and system behavior is determined by quantum laws.
We shall focus on the real part of the third-order nonlinear
dynamic susceptibility $\chi'_3(3\omega, 2\omega, \omega)$ and
denote it as $\chi'_3(\omega)$. In this paper we are interested in
the response when the $ac$ magnetic field is applied in
$z$-direction; $\chi_1=\chi_{zz}$ and so on. In formulas (2-11)
both $A_i$ and $B_i$ are the magnetic dipole moment operators.
Considering the initial value case [21] we suppose, like in [6],
that the system is in equilibrium with a small time-independent
field $h=h_z$ for $t\le 0$ and that the external field is turned
off at $t=0$, then for $t\ge 0$ the induced magnetization of the
sample in $z$-direction to first order of perturbation theory is
given by $M(t)-M_0\approx R^{(1)}(t)h$, where a relaxation
function
\begin{equation}
R^{(1)} (t)=\int_0^\infty \varphi^{(1)} (\tau)d\tau
\end{equation}          
and the first order response function is [20]
\begin{equation}
\varphi^{(1)} _{ij} (\tau)\approx -{1\over i\hbar} \langle
\left[M_i(\tau), M_j\right]\rangle \ .
\end{equation}                
The higher-order response functions are given by
\begin{eqnarray}
\varphi^{(2)} _{ijk} (\tau_1, \tau_2)&\approx &{1\over (i\hbar)^2}
\langle \left[\left[ M_i(\tau_1+\tau_2), M_j(\tau_2)\right], M_k \right]\rangle
\ ; \\                
\varphi^{(3)} _{ijkl} (\tau_1, \tau_2, \tau_3)&\approx & -{1\over
(i\hbar)^3} \langle \left[\left[\left[ M_i(\tau_1+\tau_2+\tau_3),
M_j(\tau_2+\tau_3)\right], M_k(\tau_3)\right], M_l
\right] \rangle.                 
\end{eqnarray}
The linear and the nonlinear dynamic susceptibilities (admittances
in the spectral representation [22]) may be found through response
functions (21-23). In order to find the complete expression for
susceptibility, we should use its symmetry and causality
properties [22].

The nonlinear susceptibilities may be chosen so that these
susceptibilities were symmetrical relative to simultaneous
permutation of tensor indices and corresponding to them
arguments, for example, the second rank tensors are $\chi_{ijk}
(\omega_1,\omega_2)= \chi_{ikj} (\omega_2,\omega_1)$, and the
fourth rank tensors are
\begin{equation}\chi_{ijkl}
(\omega_1,\omega_2,\omega_3) = \chi_{ikjl}
(\omega_2,\omega_1,\omega_3)= \chi_{ijlk}
(\omega_1,\omega_3,\omega_2)= \ldots \end{equation}
 according to causality property
\begin{equation}
\chi_{ij}=0\ \hbox{for} \ \tau_1<\ \hbox{max}\ (\tau_2, \tau_3,
\ldots).
\end{equation}            
Linear and nonlinear dynamic susceptibilities are given by
\begin{eqnarray}
\chi_{ij}(\omega)& =& \int_0^\infty d\tau \varphi^{(1)}_{ij}
(\tau) e^{i\omega\tau}\\        
\chi_{ijk}(\omega_1,\omega_2) &=&
{1\over 2!} \int_0^\infty d\tau_1 \int_0^\infty d\tau_2
\left\{\varphi^{(2)}_{ijk}
(\tau_1,\tau_2) e^{i(\omega_1+\omega_2)\tau_1+ i\omega_1\tau_1} \right\}\\        
\chi_{ijkl}(\omega_1,\omega_2, \omega_3) &=& {P_3\over 3!}
\int_0^\infty d\tau_1 \int_0^\infty d\tau_2 \int_0^\infty d\tau_3
\varphi^{(3)}_{ijkl} (\tau_1,\tau_2, \tau_3)\times \nonumber
\\ && \times \exp \left[i(\omega_1+\omega_2+\omega_3)\tau_1
+ i\omega_2\tau_2 + i\omega_3\tau_3 \right]        
\end{eqnarray}
In eq. (28) $P_3$ means the summation over all permutations of the
subscripts $(\omega_1j)$, $(\omega_2k)$ and $(\omega_3l)$ [22];
response functions $\varphi$ are given by expressions (21-23).

In particular, the second harmonics generation process is
characterized by tensor $\chi_{ijk}(\omega,\omega)$. If
$\omega_1=\omega_2=\omega_3=\omega$, the triple frequency
$3\omega$ is formed; the frequency tripling is described by tensor
$\chi_{ijkl} (\omega,\omega,\omega)$ (we note it as
$\chi'_3(\omega)$).

The droplet model of classical Ising spin glass was considered by
D.~S.~Fisher and D.~A.~Huse [4]. The features of this model are
described also, for example, in [23]. In the droplet model there
are only two pure thermodynamical states related to each other by
a global spin flip. In magnetic field there is no phase
transition. A droplet is an excited cluster in an ordered state
where all the spins are inverted. The natural scaling ansatz for
droplet free energy $\epsilon_L$ (which are considered to be
independent random variables) is $\epsilon_L \sim L^\theta, L\ge
\zeta (T)$; $\zeta$ is the correlation length, $L$ is the length
scale of droplet and $\theta$ is the zero temperature thermal
exponent, $\theta\le(d-1)/2$. One droplet consists of order $L^d$
spins. Below the lower critical dimension $d_l$, $\theta<0$; above
$d_l$ one has $\theta>0$.

Recently M.~J.~Thill and D.~A.~Huse [6] have shown that the
$d$-dimensional quantum ising spin glass in a transverse field
with Hamiltonian
\begin{equation}
{\cal H} = -\sum_{i,j} {\cal I} _{ij} S^z_i S^z_j - \Gamma \sum_i
S^x_i
\end{equation}
(where $S_i$ are the Pauli matrices, $\Gamma$ is the strength of
the transverse field and the nearest neighbor interactions ${\cal
I}_{ij}$ are independent random variables of mean zero) can be
represented as the Hamiltonian of the independent quantum
two-level systems (low energy droplets) of the form
\begin{equation}
{\cal H} ={1 \over 2} \sum^{\sim}_L \sum_{D_L} \left( \epsilon
_{D_L} S^z_{D_L} +\Gamma_L S^x_{D_L} \right)
\end{equation}
where $S^z_{D_L} $ and $S^x_{D_L}$ are the Pauli matrices
representing the two states of the droplet; the sum is over all
droplets $D_L$ at length scale $L$ and over all length scales
$L$, and
\begin{equation} \sum^{\sim}_L\sim \int^\infty_{L_0}
{dL\over L}
\end{equation}
with a short-distance cutoff $L_0$. The value $\Gamma_L$ regulates
the strength of quantum fluctuations ($\Gamma_L\to 0$ corresponds
to the classical limit).
\begin{equation}
\Gamma_L=\Gamma_0 e^{-\sigma L^d}
\end{equation}
is the tunneling rate for a droplet of linear size $L,\,\Gamma_0$
is the microscopic tunneling rate; $\sigma$ is defined from the
equation $2K=\sigma L^d$ where $2K$ is the surface free energy of
an interface between the two droplet states, so $\sigma$ is a
reduced surface tension for this interface; $\sigma$ is
approximately the same for all droplet. We will assume $\Gamma_L$
is the same for all droplets of scale $L$.  The droplet
excitations have a broad distribution of their free energies at
scale $L$ for large $L$ in a scaling form [4,6]
\begin{equation}
P_L (\epsilon_L) d\epsilon_L = {d\epsilon_L \over \gamma (T) L^\theta}
{\cal P} \left( \epsilon_L \over \gamma (T) L^\theta\right),\,
L\to\infty\ .
\end{equation}
It is assumed that $P_L (x\to 0) >0$, $P_L(0)-P_L(x)\sim
x^\phi$ at $x\to 0$. $\gamma(T)$ is a generalized temperature
dependent stiffness modulus which is of order of characteristic
exchange ${\cal I}= \overline{\left({\cal
I}_{ij}^2\right)}^{1\over 2} $ at $ T=0$ and vanishes for $T\ge
T_g$.

There is a crossover length scale, $L^*(T)$, defined by condition
$\Gamma_{L^*(T)}= k_BT$ or $L^*(T)=\left({1\over \sigma} \log
{\Gamma_0 \over k_B T} \right)^{1/d}$. For droplets with
$L\ll L^*(T)$ and $\Gamma_L \gg k_BT$ the energy
$\sqrt{\epsilon^2_L+\Gamma^2_L}$ is always more than $k_BT$ and
thermal fluctuations are insignificant at temperature $T$.
Droplets with $L\gg L^*(T)$ have $\Gamma_L \ll k_BT$ and
behave classically. The large droplets ($\epsilon_L\le k_B T,\,
\Gamma_L \le k_BT$) are thermally active. At low $T$ only a small
fraction of droplets is thermally active, but many low-$T$ static
properties are dominated by these droplets at the crossover
length $L^*(T)$.

The total magnet moment of a droplet will scale as $qL^{d\over 2}$
where $q$ a random number with mean zero and $\overline{q^2}\sim
q_{EA}$, $q_{EA}= \overline{\langle S^z_i\rangle^2}$ is the
Edwards-Anderson order parameter [4]. The total magnetization $M$
of the sample will be
\begin{equation}   
M=\sum^{\sim}_L {V\over L^d} \sum_{D_L} \langle S^z_{D_L} \rangle
qL^{d\over 2}
\end{equation}
and
\begin{equation}
m={M\over V}=\sum^{\sim}_L \overline{\langle S^z_{D_L} \rangle
qL^{-d\over 2}}
\end{equation}   
where $\overline{\langle S^z_{D_L} \rangle qL^{-d\over 2}}$ means
the average over the droplets energies $\epsilon_L$. The static
susceptibilities are defined in terms of the expansion of
magnetization into Taylor series approximately as
\begin{equation}
M=\chi_1 h -\chi_3 h^3 + \ldots \ .
\end{equation}         
For enough small external field it is possible to be restricted
by few terms of this expansion. The susceptibilities are got as
derivatives with respect to $h$ at $h=0: \chi_1=\left.{\partial
m\over \partial h}\right|_{h=0}$, $\chi_3= \left.{\partial^3m\over
\partial h^3}\right|_{h=0}$.

M.~J.~Thill and D.~A.~Huse have calculated static linear and
nonlinear susceptibilities for droplet system described by the
Hamiltonian (29). The static linear susceptibility diverges at
$T=0$ below the lower critical dimension $d_l$. The static
nonlinear susceptibility diverges in all dimensions $d$. The
static linear susceptibility appears to start away from the
nonzero constant $T=0$ value decreasingly versus T to lowest order
[6].

Now using the quantum droplet model of the short-range Ising spin
glass in a transverse field and quantum-mechanical case
$\beta\Gamma_L \gg 1$ (quantum regime) we calculate the third
order nonlinear dynamic response at very low finite temperatures.

When we consider the droplets at finite temperatures they may
have two characteristic rates, (a) the Rabi frequency (is of
order $\Gamma_L$) and (b) the rate of classical activation over
energy barrier B for annihilation and creation of the droplet
excitations [6]
\begin{equation}
t \sim \tau_0 e^{B\over k_BT }\ ,
\end{equation}
where $B\sim \Delta L^\psi$, $0\le\psi\le d-1$, $\psi$ is some
exponent [4], $\Delta$ is a barrier energy at $T\ll T_g$,
$\Delta\sim {\cal I}$; $\tau_0$ is a microscopic time. There is a
complicated dynamical classical-to-quantum crossover depending on
temperature, frequency of $ac$ external field and length scale
$L$. According to [6] the crossover dynamic length is determined
from the condition $\Gamma^{-1}_L= T$, i.e.
\begin{equation}
L^*_{dyn}(T) \sim \left({\sigma\over \Delta}k_B T\right) ^{1\over
\psi-d}
\end{equation}     
The system behaves presumably classically or quantum mechanically
when the dominant length scale $L$ is above or below $L^*_{dyn}$
for frequency $\omega$.

Now we consider dynamic third-order susceptibility $\chi'_3(\omega
,T)$ at finite very low temperatures (quantum regime) when
$\beta\Gamma_L\gg 1$. We define nonlinear third order dynamic
susceptibility $\chi'_3(\omega,T)$ by expression (23) and (28) $$
\chi'_3(\omega,T) = {1\over (i\hbar )^3} {P_3\over 3!}
\int_0^\infty d\tau_1 \int_0^\infty d\tau_2 \int_0^\infty d\tau_3
\exp \left[ i \left( 3\omega\tau_1 +\omega\tau_2 +
\omega\tau_3\right)\right]\times$$
\begin{equation}
\langle \left[\left[\left[ M_i(\tau_1+\tau_2+\tau_3), M_j
(\tau_2+\tau_3)\right], M_k (\tau_3)\right], M_l\right] \rangle
 .
\end{equation}                    
 The contribution of a single droplet to the
real part of dynamic third-order susceptibility up to some factor
$\sim q^2_{EA} L^{2d}$ is proportional to
\begin{equation}
\chi'_{3D_L} \sim q^2_{EA} {\left( \sum^5_{k=0} A_k
(\omega,\Gamma_L) \epsilon^{2k}_L\right) \tanh \left( {1\over
2}\beta \sqrt{\epsilon^2_L+\Gamma_L^2}\right) \over
\left(\epsilon^2_L +\Gamma_L^2\right)^{5\over 2}
\left(\epsilon^2_L +\Gamma_L^2-9\omega^2\right)
\left(\epsilon^2_L +\Gamma_L^2-\omega^2\right)^2
\left(\epsilon^2_L +\Gamma_L^2-{\omega^2\over 4}\right)\omega^2
}\ ,
\end{equation}       
where $A_0=\omega^2\Gamma^{10}_L +4\omega^4 \Gamma^8_L -5 \omega^6
\Gamma^6_L$; $A_1=2\Gamma^{10}_L - 12,5 \omega^2 \Gamma^8_L -
49,75 \omega^4 \Gamma^6_L - 23,5 \omega^6 \Gamma^4_L - 2,25
\omega^8 \Gamma^2_L $; $A_3=12 \Gamma^6_L - 45,5
\omega^2\Gamma^{4}_L +41,75 \omega^2 \Gamma^4_L$; $A_4=8
\Gamma^4_L - 15,5 \omega^2\Gamma^{2}_L$; $A_5= 2 \Gamma^2_L$.

We have to average $\chi_{3D_L}$ over droplet energies
$\epsilon_L$ using the distribution of droplet free energies (33)
and changing variables from $\epsilon_L$ to $x=\beta\epsilon_L$.

After averaging over droplet energies for cases $\Gamma_L
>3\omega, \Gamma_L \sim 3 \omega, \Gamma_L<3 \omega$ we receive
that the real part of the nonlinear susceptibility is dominated by
droplets of length scale
\begin{equation}
L_{dom(3\omega )} \sim \left(\frac{1}{\sigma}\left|\log
\left(\frac{3 \omega}{\Gamma_0}\right)\right|\right)^{\frac{1}{d}}
\end{equation}          
which is determined by condition $\Gamma_L \sim 3 \omega$. Then for
$\Gamma_L> 3 \omega$ the result of two averages is given by the following
expression
$$
\chi'_3 \sim \frac{q^2_{EA}}{\gamma^{1+\phi}} \left \{ \pi
\sec \left[\frac{\pi \phi}{2}\right]
\sum_{k=-2,0,2} A_k \omega^{k-2} \Gamma_0^{\phi -k}
\frac{(\sigma (\phi -k))^{-\alpha}}{d} {\rm G}\left[\alpha,\left|\log
\left(\frac{3 \omega}{\Gamma_0}\right)\right| (\phi -k)\right]+\right.$$
\begin{equation}\left.
\frac{\left(\frac{1}{\sigma}\left|\log
\left(\frac{3 \omega}{\Gamma_0}\right)\right|\right)^\alpha}{\alpha d}
e^{-3 \beta \omega} \sum_{k=0,1} B_k \omega^{\frac{\phi -5-2k}{2}}
\beta^{\frac{-\phi-1-2k}{2}}
\right \},
\end{equation}    
where ${\rm G}[\alpha,z]$ is incomplete gamma-function. The
coefficients in this expression depend on $\phi$ only and are
given by

$\alpha=\frac{d-\theta (1+\phi)}{d};$

$A_{-2}=-\frac{1}{4}-\frac{\phi}{4},\,A_0=-\frac{6263}{1600}-\frac{23\phi}{100}
+\frac{6441}{6400}2^{\frac{\phi+1}{2}},\,A_2=-\frac{135}{10}+12\,2^
{\frac{\phi+1}{2}},$

$B_0={\rm
G}\left[\frac{\phi-1}{2}\right]\frac{2^{\frac{\phi-7}{2}}\,3^
{\frac{\phi-1}{2}}(1435\phi^2+843\phi-1988)}{175},\, B_1={\rm
G}\left[\frac{\phi-1}{2}\right]\frac{2^{\frac{\phi-7}{2}}\,3^
{\frac{\phi-3}{2}}(791-791\phi^2)}{5}$.

$$\chi'_3(\omega,T)\sim \frac{q_{EA}^2}{\gamma^{1+\phi}}
\frac{\left(\frac{1}
{\sigma}\left|\log\left(\frac{3\omega}{\Gamma_0}\right)\right|\right)^\alpha}
{\alpha
d}\left\{C_0\pi\sec\left[\frac{\pi\phi}{2}\right]\omega^{\phi-2}+
e^{-3\beta\omega}\sum_{k=1}^4C_k\omega^{\frac{\phi+1-2k}{2}}\beta^{
\frac{-\phi+5-2k}{2}}\right\} $$ for $\Gamma_L\sim3\omega$ where

$C_0=\frac{243\,35^{\frac{\phi-1}{2}}2^{7-\phi}}{185}-\frac{68931\,2^{3\phi-3}}
{925}+\frac{3^\phi}{2}-\frac{391\,2^{\frac{\phi+5}{2}}3^\phi}{925}-\frac{27\,
2^{\frac{3\phi-5}{2}}\phi}{5},\,C_1=-{\rm
G}\left[\frac{\phi-1}{2}\right]\frac{2^{
\frac{\phi-3}{2}}\,3^{\frac{\phi+3}{2}}}{5},$

$C_2={\rm G}\left[\frac{\phi-1}{2}\right]\frac{2^{
\frac{\phi-3}{2}}\,3^{\frac{\phi+1}{2}}}{5}-{\rm
G}\left[\frac{\phi+1}{2}
\right]\frac{2^{\frac{\phi+1}{2}}\,3^{\frac{\phi-5}{2}}(458+945\phi)}{35},$

$
C_3=-{\rm G}\left[\frac{\phi+1}{2}
\right]\frac{11897\,2^{\frac{\phi-7}{2}}\,3^{\frac{\phi-5}{2}}}{175}+
{\rm G}
\left[\frac{\phi+3}{2}\right]\frac{10206\,2^{\frac{\phi-7}{2}}\,3^{
\frac{\phi-5}{2}}}{35},$

$
C_4=-{\rm G}
\left[\frac{\phi+3}{2}\right]\frac{359106\,2^{\frac{\phi-7}{2}}\,3^{
\frac{\phi-5}{2}}}{175}$.

This expression does not diverge if $2<\phi<3$ and
$-1+\frac{d}{\theta}<\phi$. This expression has singularity at
$\omega\sim \frac{\Gamma_0}{3}$. When the frequency increases the
values of $\chi'_3$ are growing to infinity while
$\omega\to\frac{\Gamma_0}{3}$. $\chi'_3$ maintains this property
when $\omega$ is more than $\frac{\Gamma_0}{3}$. The temperature
dependence of $\chi'_3$ in this case has no extremes, we observe
monotonous decrease of values of $\chi'_3$ with temperature.

 Nonlinear susceptibility given by expression (42) does
not diverge if $2<\phi<3$ and $-1+\frac{d}{\theta}<\phi$.

We observe that nonlinear susceptibility has strong dependence on
distribution function $P_L(\epsilon_L)$, i. e. on $\phi$, on
droplet microscopic tunneling rate $\Gamma_0$ and other
parameters. One can see that the real part $\chi'_3(\omega,T)$
varies approximately logarithmically with frequency. This
signalizes broad distribution of relaxation times of the system.

Let us take for numerical calculation the following numbers:
$\theta=1,\Gamma_0=10^{10}s^{-1},d=3,\phi=2.5,\gamma=10^{-15}erg,
\sigma=10^{-15}, q=0.5$. The frequency dependence of $\chi'_3(\omega,T)$
is shown in Fig. 1. We give $\log_{10} f$-dependence at $\log_{10} f$
from 0 to 11 at fixed several temperatures:
$T_1=0.001,T_2=0.005,T_3=0.01,T_4=0.05$.

The frequency interval covers some decades of frequencies. Our numerical
calculations show the crossover between low-$\omega$ and high-$\omega$
behaviors. In low-$\omega$ region the nonlinear response is found nonsingular
and slowly decreasing. When frequency increases the curve falls down more
quickly, the nonlinear response diverges at $\omega \sim \frac{\Gamma_0}{3}$,
then the curve rises to some value. In low-$\omega$ region we have a
qualitative agreement with experimental data for disordered dipolar magnet
${\rm LiHo}_x{\rm Y}_{1-x}{\rm F}_4$. At different low fixed temperatures
the behavior of $\chi'_3(\omega,T)$ is the same and the values of
$\chi'_3(\omega,T)$ are approximately the same. Therefore we give only one
curve for all fixed temperatures.

In Fig. 2 we give the temperature dependence of
$\chi'_3(\omega,T)$ at temperatures from 0 to $10^{-2}$K at fixed
several frequencies: $f_1=10^7Hz, f_2=2.5\times 10^7Hz,f_3=5
\times 10^7Hz,f_4=7.5\times 10^7Hz,f_5=10^8Hz$ of $ac$ field ($
f=\frac{\omega}{2\pi}$). The behavior of $\chi'_3(\omega,T)$
indicates the following glassy-like features. The curves of the
temperature dependence of $\chi'_3(\omega,T)$ have maxima
depending on fixed frequency. The temperature of $\chi'_3$-maximum
$T_f(\omega)$ depends on frequency. The nonlinear susceptibility
magnitudes at different fixed frequencies are remarkably
distinguishable. The temperatures of maximum values are different.
When the frequency increases the temperature of $\chi'_3$-maximum
shifts towards high temperatures. The similar curve of temperature
variation of $\chi'_3$ was observed in spin glasses at more high
temperatures [2,13]. If we consider only T-dependent part of
$\chi'_3$ we see that the $\chi'_3$-maxima are sharp (Fig. 3).

The cubic dynamic susceptibility $\chi'_3(3\omega)$ is
analitically and numerically calculated in quantum spin glass in
terms of quantum droplet model on the basis of general dynamic
nonlinear quantum-mechanical response theory. We have carefully
analyzed the susceptibility temperature-frequency behavior to
study the properties of the low temperature magnetic state and to
determine whether or not a conventional spin glass state exists
below $T_f$. Comparing with the case of a true spin glass
transition we see that our data indicate that the magnetic state
below $T_f$ does not correspond to a conventional spin glass state
below $T_f$. We find a glassy type slow dynamics. Similar
frequency dependence was observed by W. Wu et al.[14].

Our calculations at $T=0$ coincide with $T=0$ result of
M.~J.~Thill and D.~A.~Huse [6]. For finite temperatures we find
some features which have been recently observed [13]. We suppose
that at some very low temperature $T_f$ (temperature of maximum of
$\chi'_3(\omega,T)$) there is a phase transition. If $\theta>0$
and $d=3$ we suppose a true phase transition at very low
temperature $T_f\sim 10^{-4}\div8.5\times 10^{-4}K$ for $f=10^7
\div10^8Hz$ respectively (Fig. 3).

Besides frequency and temperature dependence the shape of
$\chi'_3(3\omega)$ depends crucially on the probability
distribution of droplet free energies, on the tunneling rate for a
droplet of linear size L, on the material parameters. In
consequence of this dependence there is divergence (or
convergence) of $\chi'_3(3\omega)$. We need to take into account
(in future paper) the dipole-dipole interaction between droplets
and also droplet-lattice interaction.

Applying our results to the reported experimental data on the
nonlinear dynamic susceptibility of ${\rm LiHo}_x{\rm Y}_{1-x}{\rm
F}_4$ we observe that a fairly good agreement may be achieved.

\newpage
\centerline{\Large References}

\begin{enumerate}
\item Glassy dynamics and optimization (Ed.: Y.L. van Hemmen, M.
Morgenstern) Springer Verlag, Berlin Heidelberg, 1987.

\item M.Mezard, G.Parisi and M.A.Virasoro "Spin Glass Theory and
Beyond",World Scientific, Singapore, (1987);
 A.P. Young (Ed.) "Spin-glasses and random fields", World Scientific, Singapore, (1997).

\item T. Kopec, Phys. Rev. Lett. {\bf 79}, 4266 (1997)

\item D. S. Fisher and D. A. Huse, Phys. Rev. Lett. {\bf 56}, 1601 (1986); Phys. Rev. B
{\bf 36}, 8937 (1987); Phys. Rev. B {\bf 38}, 373,386 (1988);
D.S. Fisher, J. Appl. Phys. {\bf 61}, 3672 (1987).

\item A. Barrat and L. Berthier, preprint cond-mat/0102151
8 Feb. 2001;

Y.G. Joh and R. Orbach. Phys. Rev. Lett. {\bf 77}, 4648 (1996).

\item M.J. Thill and D.A. Huse,  Physica {\bf A241} , 321 (1995)

\item H. Ishii and T. Yamamoto, J. Phys.  {\bf C18}, 6225 (1985);

\item N. Read, S. Sachdev and J. Ye, Phys. Rev. B{\bf 52}, 384
(1995).

\item H. Rieger and A.P. Young, Phys. Rev. Lett. {\bf 72}, 4141 (1994).

\item L.P. L\'evi and A.T. Ogielski, Phys. Rev. Lett {\bf 26},
3288 (1986)

\item L.P. L\'evi, Phys. Rev. Lett B{\bf 38},4963 (1988)

\item T. Jonsson, K. Jonason, P. J\"onsson and P. Nordblad, Phys. Rev. B{\bf 59}, 8770 (1999)

\item K. Gunnarsson et al. Phys. Rev. B. {43}, 8199 (1991);

M. Hagiwara et al. J MMM, {\bf 177-181}, 175 (1998).

\item W. Wu, D. Bitko, T. F. Rosenbaum and G. Aeppli, Phys. Rev. Lett.
{\bf 71}, 1919 (1993);

 D. Bitko, T. F. Rosenbaum and G. Aeppli, Phys. Rev. Lett.
{\bf 75}, 1679 (1995).

\item J. Mattsson, Phys. Rev. Lett. {\bf 75}, 1678 (1995).

\item G. Busiello and R. V. Saburova, Int. J. Mod.  Phys. {\bf B14}, 1843 (2000);

R. Saburova, G.P. Chugunova and G. Busiello. The physics of metals
and metallography, {\bf 87}, 509-515 (1999);G. Busiello, R. V.
Saburova and V.G. Sushkova, Sol.State Comm.{\bf 119},545 (2001).

\item R. Kubo and K. Tomita, J. Phys. Soc. Japan {\bf 9}, 888 (1954)

R. Kubo , J. Phys. Soc. Japan {\bf 12}, 570 (1957)

\item W. Bernard and H.B. Callen, Rev. Mod. Phys. {\bf 31}, 1017
(1959).

R.L. Peterson, Rev. Mod. Physics {\bf 39}, 69 (1967)

 \item W.T. Grandy, Jr. "Foundation of statistical mechanics" D.
 Reidel Publishing Company, Dordrecht, Holland, 1988.

 \item W. Brenig "Statistical theory of heat" Springer-Verlag,
 1989.

 \item R.L. Stratonovich "Nonlinear nonequilibrium thermodynamics
 I" Springer-Verlag, 1992;

 V.M. Fain. Kvantovaya Radiophizika. Izdatelstvo Sovietskoe Radio.
 Moskva, 1972 (Russian).

 \item J.A. Mydosh "Spin Glasses: an experimental introduction.
 Taylor \& Francis, London, 1993.

 \item P. Esquinazi (Ed.) Tunneling systems in amorphous and
 crystalline solids. Springer-Verlag-Heidelberg, 1998.

 \end{enumerate}
 \newpage
\centerline{\Large Figure captions}

Fig.1 - The frequency dependence of the real part of the nonlinear
dynamical susceptibility at fixed temperature.

Fig.2 - The temperature dependence of the real part of the
nonlinear dynamical susceptibility at various frequencies.

Fig.3 - The temperature dependence of the T-dependent part of the
  nonlinear dynamical susceptibility at various frequencies f.

 \end{document}